\documentclass{aastex}
\usepackage{natbib}
\usepackage{longtable}
\usepackage{subfigure}
\usepackage{lscape}
\usepackage{graphicx}
\usepackage{multirow}
\usepackage{float}
\usepackage{color}
\usepackage{bm}
\newcommand       \mum          {\,{\rm \mu m}}
\newcommand       \Teff         {T_\mathrm{eff}}
\newcommand       \Ks           {K_S}

\begin{document}
\title{Revision of Stellar Intrinsic Colors in the Infrared by the Spectroscopic Surveys}
\author{Mingjie Jian\altaffilmark{1},
		Shuang Gao\altaffilmark{1,*},
        He Zhao\altaffilmark{1},
		Biwei Jiang\altaffilmark{1}}

\altaffiltext{1}{Department of Astronomy,
                 Beijing Normal University,
                 Beijing 100875, China;
                 \tt mkk@mail.bnu.edu.cn,
                 \tt hezhao@mail.bnu.edu.cn,
                 \tt bjiang@bnu.edu.cn}

\altaffiltext{*}{Corresponding author: {\tt sgao@bnu.edu.cn}}

\begin{abstract}
Intrinsic colors of normal stars are derived in the popularly used infrared bands involving the 2MASS/$JH\Ks$, \emph{WISE}, \emph{Spitzer}/IRAC and \emph{AKARI}/S9W filters. Based on three spectroscopic surveys -- LAMOST, RAVE and APOGEE, stars are classified into groups of  giants and dwarfs, as well as metal-normal and metal-poor stars. An empirical analytical relation of the intrinsic color is obtained with stellar effective temperature $\Teff$ for each group of stars after the zero-reddening stars are selected from the blue edge in the $J-\lambda$ versus $\Teff$ diagram. It is found that metallicity has little effect on the infrared colors. In the near-infrared bands, our results agree with previous work. In addition, the color indexes $H-W2$ and $\Ks-W1$ that are taken as constant to calculate interstellar extinction are discussed. The intrinsic color of M-type stars are derived separately due to lack of accurate measurement of their effective temperature.
\end{abstract}

\section{Introduction}

Stellar intrinsic color, or color index, is a fundamental parameter associated with the properties of stellar atmosphere, which also reveals the information about spectral energy distribution, i.e. bolometric corrections \citep{Lee1970}. Intrinsic color is  of vital importance in estimating color excess and extinction$/$extinction law, see e.g. \citet{Xue2016}. The intrinsic colors of visible bands are well determined by \citet{visibleic} and by stellar models. However, due to numerous molecular absorption bands in the infrared, theoretical determination of stellar color indexes suffers some uncertainties. In addition, space infrared astronomy develops non-traditional filters for intended scientific goals. The intrinsic color indexes have not been systematically studied for these newly developed filter bands.

\citet{Johnson} derived the first widely used intrinsic colors in $UBVRIJHKLMN$, among which $JHKLMN$ are classical infrared bands in accordance with the terrestrial atmospheric windows. He assumed that there is no interstellar reddening within 100 pc from the Sun, so the average of the observed colors of such nearby stars was taken as intrinsic for dwarfs and giants of various spectral types. This must have over-estimated the intrinsic colors because these nearby stars should suffer some (even though small) interstellar extinction. Following the pioneering work of \citet{Johnson}, \citet{Lee1970}, \citet{Bessell1988} and \citet{Bouchet1991} determined intrinsic color indexes in the same way. However, \citet{Kooenneef1983} pointed out that nearly all the results obtained before 1983 needed to be modified for comparison because of slight difference in photometric system they used. Even with modification, variety in method of de-reddening, for example, using ``mean'' extinction law to all stars, or even neglecting interstellar reddening (see the description of \citealt{dougherty1993}), brings about dispersion of the results and makes the estimation of uncertainty difficult \citep{wegner1994}.

\citet{Ducati} invented a new method to determine stellar intrinsic color in the traditional infrared Johnson system. Based on a large catalog of infrared photometric observations (CIO catalog; \citealt{CIO}) of stars whose optical spectral types are identified, a zero-reddening curve is delineated by the blue envelope in the diagram of color index $V-\lambda$ versus  effective temperature $\Teff$. The advantage of this method over \citet{Johnson} is that the zero-reddening stars are searched in a type of stars for the bluest color, which avoids inclusion of interstellar extinction in color index. However,  the catalog (3946 sources) used by \citet{Ducati} has no complete sample of every sub-type of stars from B0 to M4 dwarfs. Consequently, resultant intrinsic color indexes only roughly resolve the spectral types. In addition, their determination of the blue envelop in the  $V-\lambda$ vs. $\Teff$  diagram looked a bit arbitrary. Nevertheless, the Ducati method can in principle lead to very precise determination of intrinsic color indexes if stellar parameters of enough large sample of stars are measured for a given type of star.

Since the work of \citet{Ducati}, large-scale spectroscopic surveys of stars have been carried out. RAVE (the RAdial Velocity Experiment) started in 2003, and until 2013 (end of observation campaign), has obtained 574,630 spectra of 483,330 unique stars in the magnitude range 8 $< I <$ 12\,mag \citep{Ko}. The Large Sky Area Multi-Object Fiber Spectroscopic Telescope (LAMOST) (\citealt{LAMOST} and \citealt{LEGUE}) started its observation in 2011, and has acquired the stellar parameters of more than two million stars until the second data release in 2015 \citep{LAMOSTP, LAMOSTDR1}. APOGEE (The APO Galactic Evolution Experiment, \citealt{APOGEE}) obtained high-resolution H-band spectra of more than 100,000 giants. All these spectroscopic surveys calculated basic stellar parameters ($\Teff$, $\log{g}$ and [Fe/H]), which provide a fantastic database for studying the intrinsic colors. On the other hand, infrared photometric surveys also enlarged greatly the sample for selecting zero-reddening stars. Ground based 2MASS survey was an unbiased all-sky survey in the near-infrared bands, with $JH$ in accordance with the Johnson system, and a new $\Ks$ band which has a short cut at the long wavelength end. The space projects, \emph{Spitzer}, \emph{WISE} and \emph{AKARI}, observed either all the sky or a large part of the sky. They adopted completely new filter bands in mid-infrared in which no intrinsic color indexes have been determined for stars.

In this work we apply the basic idea of the method of \citet{Ducati} to the greatly improved large-scale photometric and spectroscopic data to determine the infrared intrinsic color indexes of normal stars. We first describe the data in Sec.~\ref{sec:data}, and then the method in Sec.~\ref{sec:method}. The result and discussion are presented in Sec.~\ref{sec:result}, and finally we summarise our work in Sec.~\ref{sec:summary}.

\section{Data and Sample Selection}
\label{sec:data}

The filter bands to deal with involve the most popularly used, that is, in the large-scale survey. Besides, these bands are different from the Johnson system. Specifically, we study the intrinsic colors of stars related to the 2MASS/$JH\Ks$ in the near-infrared, the \emph{Spitzer}/IRAC, \emph{WISE} and \emph{AKARI}/S9W bands in the mid-infrared. The 2MASS/$JH$ bands conform to the Johnson system and form the bridge to compare with classical results.

\subsection{Photometric and Spectroscopic Data}

Both photometric and spectroscopic data are taken from a few surveys, which expand the band coverage and enlarge the samples.

\subsubsection{Photometric data}
For the $J$ (1.25\,$\mum$), $H$ (1.65\,$\mum$) and $\Ks$ (2.17\,$\mum$) bands, the data are taken from the Two Micron All Sky Survey (2MASS, \citealt{Sk}). It used two highly automated telescopes operated between 1997 and 2001, provided a huge amount of near infrared photometries covering the whole sky. The limiting magnitudes of point source are up to 15.8, 15.1 and 14.3 (with signal-noise ratio S/N greater than 10), in the $J$, $H$ and $\Ks$ bands respectively.

The Wide-field Infrared Survey Explorer (\emph{WISE}) has four bands, i.e. $W1$ ($\lambda_{\rm eff}=$ 3.35\,$\mum$), $W2$ (4.60\,$\mum$), $W3$ (11.56\,$\mum$) and $W4$ (22.08\,$\mum$) band with a bandwidth of 0.66, 1.04, 5.51 and 4.10\,$\mum$ respectively. \emph{WISE} began to map the sky after its launch in 2009 \citep{WISE} and covered most of the sky area more than eight times. After  depletion of secondary cryogen tank, the first two bands ($W1$ and $W2$) continued to operate, finishing the survey called NEOWISE Post-Cryogenic Mission \citep{NEOWISE}. \emph{WISE} achieved a limiting magnitude of 16.9, 16.0, 11.5 and 8.0\,mag at 5$\sigma$ level in $W1$, $W2$, $W3$ and $W4$.  Combining data from WISE All-Sky \citep{AllWISE} and NEOWISE, a more precise and comprehensive catalog, ALLWISE, is obtained in these four bands.

The Galactic Legacy Infrared Mid-Plane Survey Extraordinaire (GLIMPSE, \citealt{GLIMPSE1, GLIMPSE2}) is a survey carried out by the \emph{Spitzer} satellite in four bands of Infrared Array Camera (IRAC). These bands are usually designated as [3.6], [4.5], [5.8] and [8.0] with $\lambda_{\rm eff}$ at 3.55, 4.49, 5.73 and 7.87\,$\mum$ and a bandwidth of 0.75, 1.01, 1.43 and 2.91\,$\mum$ respectively. The \emph{Spitzer}/IRAC [3.6] and [4.5] bands are thus very similar to the \emph{WISE}/$W1$ and $W2$ bands in both effective wavelength and bandwidth.

\emph{AKARI} (Astro-F, \citealt{AKARI}) is a Japanese infrared satellite aiming to provide a survey with a higher precision than IRAS. Operating from 2006 to 2011, \emph{AKARI} surveyed all sky in the S9W and L18W bands using camera MIR-S and MIR-L. The \emph{AKARI}/S9W and L18W bands have an effective wavelength $\lambda_{\rm eff}$ of 8.23 and 17.61\,$\mum$ and a bandwidth of 4.10 and 9.97\,$\mum$, which were dedicated to the silicate features at 10 and 20\,$\mum$.

\subsubsection{Spectroscopic data}

In the last decade, the advent of multi-fiber spectrographs makes it feasible to obtain very large amount of spectra. For the purpose of determining stellar intrinsic colors in the infrared, the stellar parameters are adopted from three large-scale spectroscopic surveys, specifically, the RAVE, LAMOST, and APOGEE survey.

RAVE \citep{Ko} is a multi-fiber stellar spectroscopic astronomical survey in the Australian Astronomical Observatory (AAO), which targets mainly the southern sky. In addition to the radial velocity (RV) as the most important parameter of this project, RAVE also obtained fundamental stellar atmospheric parameters: effective temperature $\Teff$, surface gravity $\log{g}$, and metallicity [Fe/H]. Recent DR4 includes 425,561 stars. The stellar position is matched to the 2MASS Point Source Catalog (PSC), which brings great convenience to associate stellar parameters with the observed colors.

LAMOST, with 4000-fibers, is a meridian reflecting Schmidt telescope of National Astronomical Observatories of China. The ``LAMOST Experiment for Galactic Understanding and Evolution" or LEGUE began in 2011 aiming to survey the whole Milky Way in north celestial sphere inaccessible by RAVE. The DR2 (second data release) catalog contains stellar radial velocity (RV), $\Teff$, $\log{g}$, and [Fe/H].  The range of $\Teff$ is from 3500\,K to 8500\,K.

Both RAVE and LAMOST/LEGUE projects observed more dwarf stars than giants in optical. Differently, APOGEE \citep{APOGEE} targeted intentionally giant stars in the infrared. APOGEE uses high-resolution, high S/N spectroscopy around $H$ band to penetrate the dust that obscures significant fractions of the disk and bulge of our Galaxy \citep{Ni}. Over 100,000 giant stars are observed across the Galactic bulge, halo,  and disk. The first APOGEE data release is part of SDSS/DR12, which includes also the basic stellar parameters, $\Teff$, $\log{g}$, [Fe/H] etc. The survey excluded most dwarfs, and its $H$ band magnitude limit of spectra is up to 12.2.

The M-type sample is selected independently. Stellar parameters are reliable for stars in the middle spectral types, such as F-, G- and K-type, but very uncertain for M-type stars due to numerous molecular absorption bands which lead to poor match with spectral templates. Specific methods are developed to classify M-type stars separately. For M-type star sample, we choose the results of \citet{MTypeD} for dwarfs from the LAMOST Pilot Survey and \citet{MTypeG} for giants from the LAMOST DR1. Though the scale of these dataset is large (67,082 dwarfs and 10,044 giants), they only have a small fraction (less than 20\%) of stars overlapping with LAMOST/DR2. The M-type giant, as pointed out in \citet{MTypeG}, has 4.7\% of dwarf contamination in it. According to the color difference in $J-\Ks$ and $W1-W2$ between M-type dwarfs and giants \citep{MTypeG}, the criteria $J-\Ks>0.8$ and $W1-W2<-0.1$ are applied to the M-type giant sample in order to remove the contamination of M-type dwarfs.  Because no stellar parameters are available, no exclusion is made to the M-type catalogs, and only photometric quality is controlled in the counterparts.

\subsection{Cross-Identification and Reduction}

The photometric and spectroscopic catalogs are cross-identified by positional matching within 3$\arcsec$. This cross identification associates apparent brightness in infrared bands with stellar atmospheric parameters which are derived from either optical or infrared spectroscopy. The quality of stellar parameters is controlled by the claimed errors. Because each survey has its own accuracy, criterion changes depending on the survey. Table~\ref{tab:spec-restrict} lists the details of data quality control for three spectroscopic surveys. The primary parameter $\Teff$ that decides the color is required to be more accurate than 200\,K. On error of $\log g$, the LAMOST catalog covers a wide range from 0.15 to 1.0\,dex, which forces us to set a very loose constraint, 0.7\,dex, in order to keep a big enough sample. Fortunately, $\log{g}$ is a secondary factor to influence the color as far as stellar luminosity class is correctly determined.

The photometric error is constrained depending on the quality of the survey. Table~\ref{tab:phot-restrict} lists the quality of photometry in corresponding bands. For an accurate determination of the intrinsic colors, we need both a high accuracy of photometry and a large sample of stars. Taking these two factors into account, the error in the band with large amount of stars is more strictly constrained at 0.05\,mag or less in the 2MASS bands, the \emph{Spitzer}/IRAC bands and the first three\emph{ WISE} bands. Meanwhile, the constraint is relaxed to 0.1\,mag in the $W4$ and S9W bands due to their much smaller sample of stars. The sample of cross-identification is presented in Table~\ref{tab:phot-restrict}.

The cross-identification results in very limited number (844 with LAMOST, 1624 with APOGEE and 238 with M-type giant catalog) of stars for the \emph{AKARI}/S9W band. This can be understood by the relatively low sensitivity of S9W band, which is 7.6\,mag at a 5$\sigma$ level, and the infrared-bright stars are not necessarily bright in optical. Moreover, only 20 stars in LAMOST and 100 stars in APOGEE have counterparts in the \emph{AKARI}/L18W catalog so the L18W band is not taken into account for further analysis. The catalog of \emph{Spitzer}/GLIMPSE provides a good cross-identification with APOGEE but only GLIMPSE 360 has overlapping area with LAMOST, thus the LAMOST stars have only IRAC photometry in the [3.6] and [4.5] bands.

\subsection{Classification of stars: $\log{g}$ and \rm{[Fe/H]}}

\begin{figure}[H]
\figurenum{1}
\plotone{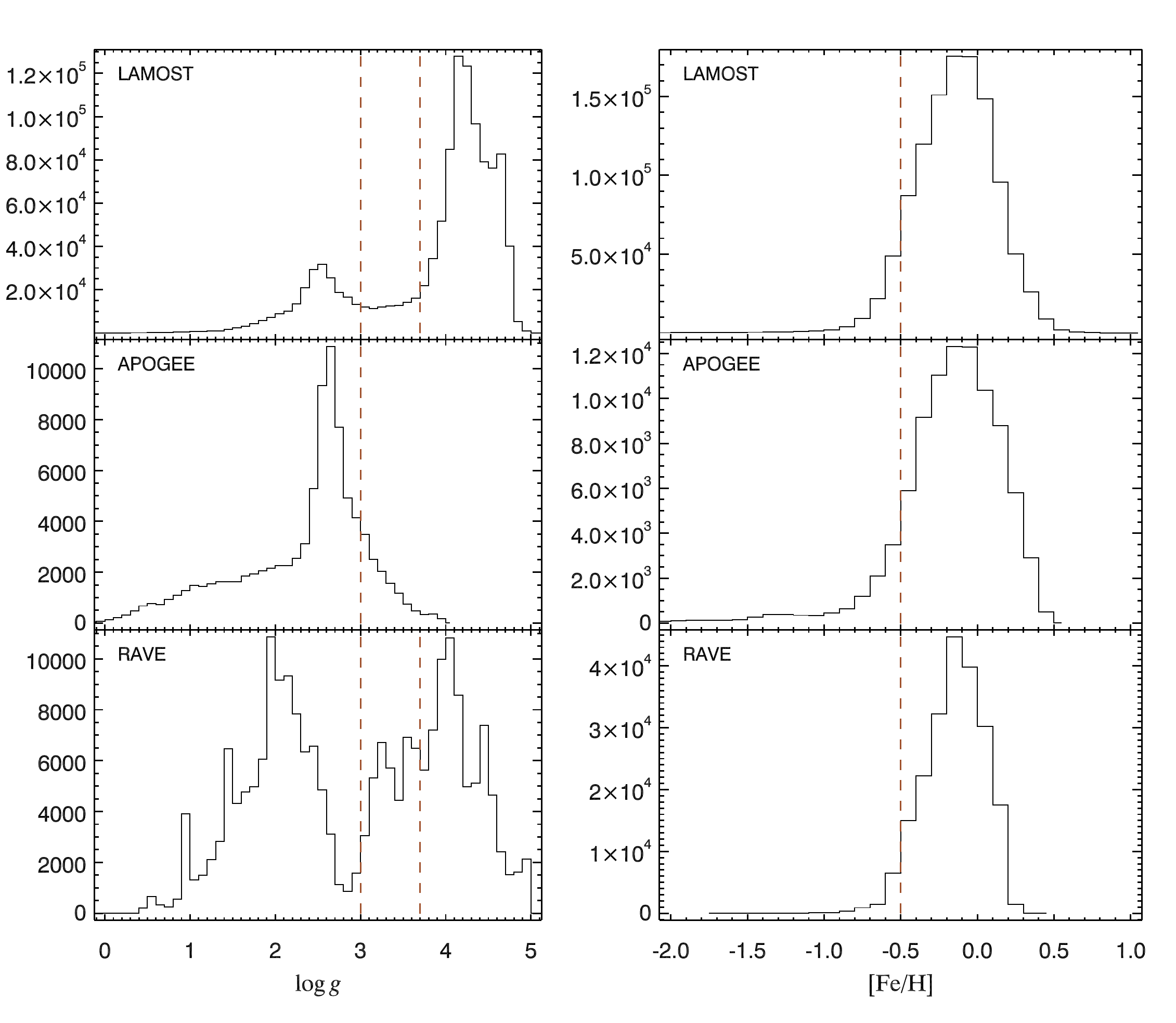}
\caption{Distribution of $\log{g}$ and [Fe/H] from spectroscopic surveys. Brown dashed lines are borders used to separate dwarf and giant  stars.}
\label{fig:hist}
\end{figure}

Apart from $\Teff$, $\log{g}$ and [Fe/H] also play some roles in affecting color index, but they are secondary factors. The stars are classified roughly into giant and dwarf according to $\log{g}$, and into metal-normal and metal-poor stars according to [Fe/H]. Fig.~\ref{fig:hist} shows the distribution of $\log{g}$ and [Fe/H] of stars in the spectroscopic surveys after cross identification with the 2MASS survey.  There is a clear double-peak distribution in the $\log{g}$ histogram distinguishing dwarf and giant in both LAMOST and RAVE. Due to the selection bias of APOGEE to giants, few stars have $\log{g}$ larger than 4.0. \citet{logg} set a value of $\log{g}=3.5$ as the boundary of giant and dwarf. Taking the average $\log{g}$ uncertainty of LAMOST (0.47\,dex) and RAVE (0.16\,dex) into account,  we shift the boundary leftward and rightward to classify stars with $\log{g}<3.0$ as ``giant'' and $\log{g} > 3.7$ as ``dwarf''. The stars with $3.0 < \log{g} < 3.7$ are dropped to avoid ambiguity. For the APOGEE database, only giants are picked up and the rest is excluded.

Metallicity has much weaker influence on infrared colors than in optical \citep{metalop}. The modelling of metal-poor stars usually has relatively higher uncertainty than metal-normal stars. The stars are simply classified into metal-normal and metal-poor with a boundary at [Fe/H]=$-0.5$. The numbers of each class in three surveys are listed in Table~\ref{tab:class}.

\section{Method: the Blue Edge}
\label{sec:method}

We adopted the basic idea of the method of \citet{Ducati} to determine the intrinsic color indexes: the bluest star for a given spectral type has the smallest interstellar extinction. In case that the sample includes star with no interstellar extinction, the observed color of the bluest star is equivalent to the intrinsic color of the given type. \citet{WJ2014}  made use of this method successfully to study the near-infrared extinction law. With the support of above-mentioned large-scale photometric and spectroscopic surveys, our method has two advantages over the \citet{Ducati} method: one is that the bluest stars will be determined mathematically, which avoids arbitrary uncertainty; the other is that an empirical analytical relation of color index to effective temperature will be derived so that the intrinsic color indexes can be calculated conveniently from effective temperature. The key point in this method is then to define a blue edge in the color -- $\Teff$ diagram. How to define the blue edge is described as follows.

\begin{figure}[H]
\figurenum{2}
\plotone{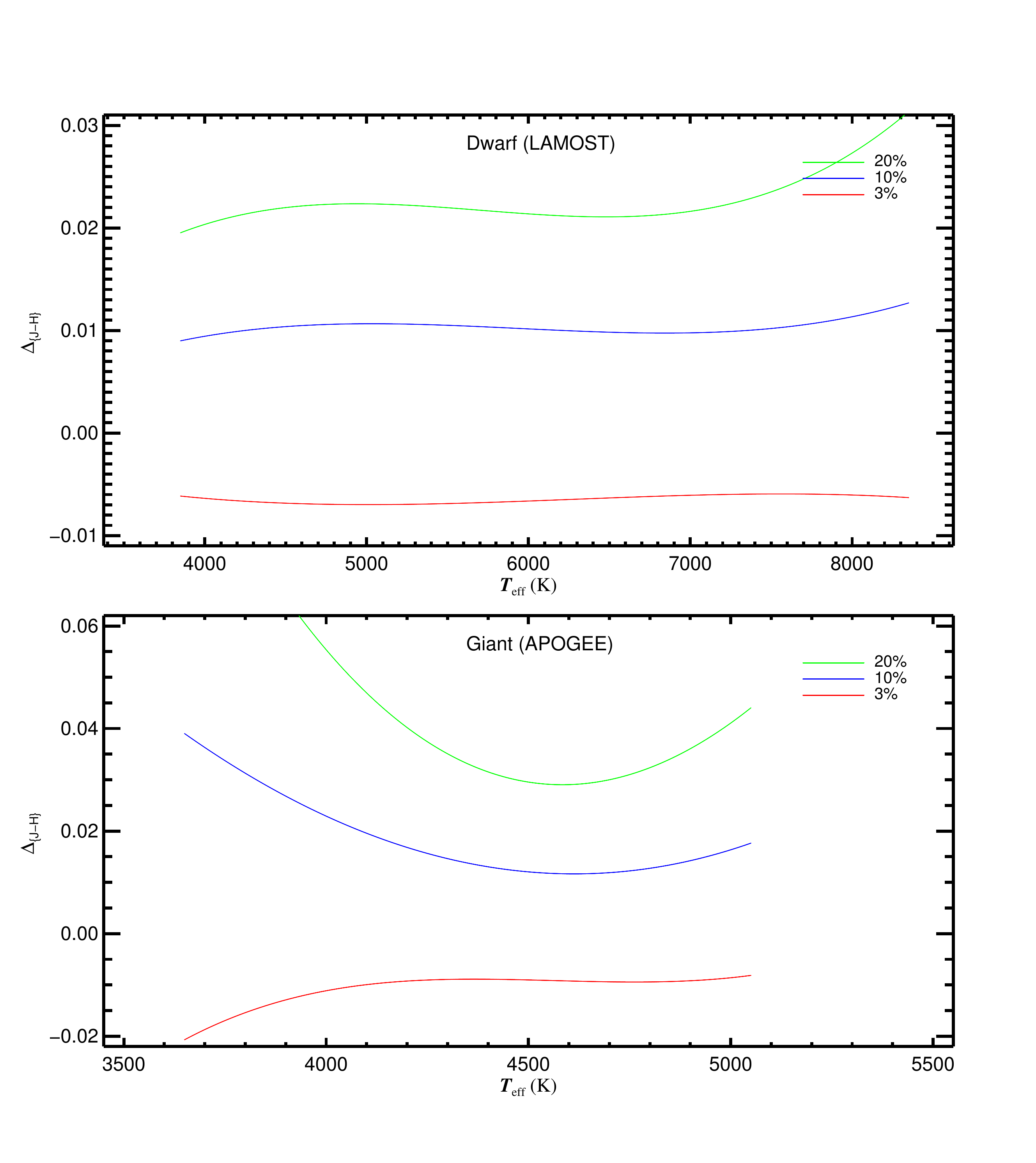}
\caption{The difference in J-H of three percentages in choosing the bluest stars with the 5\% result.}
\label{fig:3510}
\end{figure}

\begin{enumerate}
\item The whole sample in the color -- $\Teff$  diagram is divided into different bins according to $\Teff$. The final adopted bin size is 50\,K. An appropriate interval of $\Teff$ should reflect the average error of $\Teff$ ($\sigma_{\Teff}$).  The $\sigma_{\Teff}$ distributes from several tens to hundreds Kelvins in the LAMOST official data with a typical value of about 100\,K to 150\,K. But an internal uncertainty of 50\,K to 100\,K is obtained for spectra with S/N $>$ 20 by comparing two-epoch observations of the same stars \citep{Gao}. Moreover, when the bin size of $\Teff$ changes from 50\,K to 137\,K (the mean error of $\Teff$ of the selected LAMOST stars), the difference in the resultant color indexes is on the order of 0.01~mag (0.012 at $J-H$), smaller than the photometric uncertainty, which means the bin size ranging from 50\,K to 150\,K has no significant effect on the result. Thus, a bin size of 100\,K of $\Teff$ is adopted for the colors $J-W4$, $J-\mathrm{IRAC}$, $J-\mathrm{S9W}$ of the APOGEE stars, and $H-[4.5]$ for both the APOGEE and LAMOST stars, in order to guarantee enough number of sources in each bin for further fitting.
\item The bins with number of stars less than certain values are considered having no clear edge and are excluded. The cutting value of number of stars is 100 for $J-H$, $J-\Ks$, $J-W1$ and $J-W2$, 50 for $J-W3$ and $J-W4$ (LAMOST), 10 for $J-W4$ (APOGEE), $J-\mathrm{IRAC}$ bands, $J-\mathrm{S9W}$ and all bands of M-type. The numbers are adjusted according to the scale of the sample that a larger sample has a higher cutting value.
\item The stars with a color more than 3$\sigma$ outlying to the mean value of a bin are excluded. But the $J-W4$ and $J-\mathrm{IRAC}$ colors for LAMOST stars are exceptions, a 2$\sigma$ rule replaces  because the variance is large.
\item Indeed, the bluest star in each $\Teff$ bin is not taken as the standard of intrinsic color of this $\Teff$. Because of the photometric error, the bluest star with the photometric error should be bluer than the intrinsic color in the absence of interstellar extinction. Instead, the median color of the 5\% bluest stars is taken as the intrinsic color of this $\Teff$ bin. This leaves some stars bluer than the assigned intrinsic color, but mostly within a distance less than the photometric error. It is very difficult to make the choice of the bluest fraction (i.e. 3\%, 5\%, 10\% and 20\%), as higher fraction will shift upward the intrinsic color line. Various ways are tried to select the zero-reddening stars, e.g. whether the residual distribution is Gaussian, or the residual mean matches the photometric error. No way shows clear sign of the borderline for zero-reddening stars. On the other hand, the increase of the expected line of intrinsic color is very small when the percentage increases from 3\% to 20\%.  Fig.~\ref{fig:3510} indicates that the difference in $J-H$  is within 0.02 with a choice of 3\%, 5\% to 10\% bluest in $J-H$, which matches the photometric error. Besides, \citet{Xue2016} and \citet{WJ2015} both adopted 5\% in their study, and the result after 5\% is consistent with \citet{Bessell1988}. Therefore, we follow previous studies to choose the 5\% bluest stars, although there is no solid mathematical proof.
\end{enumerate}

\begin{figure}[H]
\figurenum{3}
\plotone{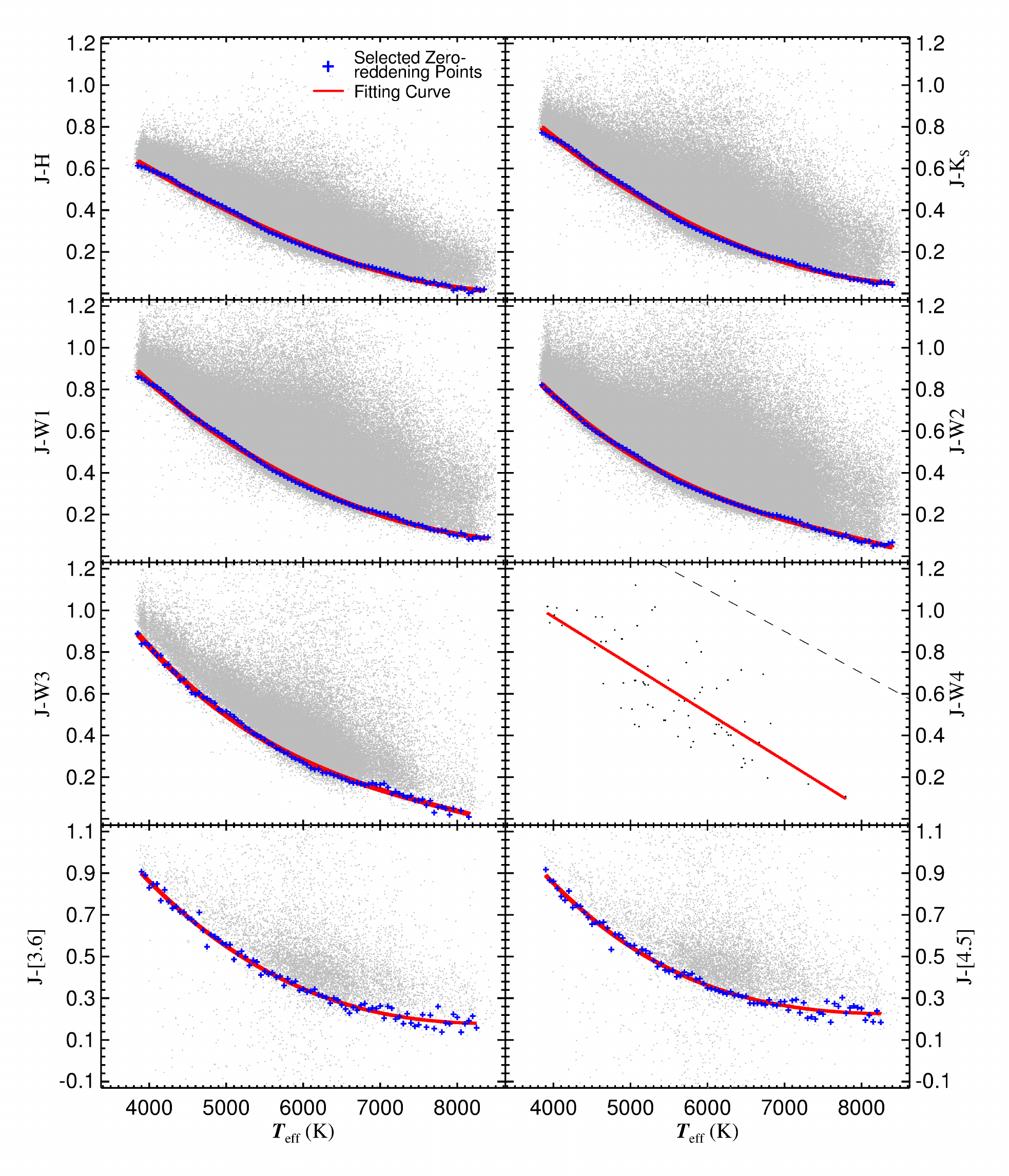}
\caption{Color -- $\Teff$ diagrams of dwarf stars from the LAMOST survey. Gray/black points decode all the stars that pass data quality control, blue crosses decode the selected zero-reddening stars and red line is the fitting curve. The stars under the dashed line in $J-W4$ diagram is adopted for a linear fitting.}
\label{fig:ic-d}
\end{figure}

\begin{figure}[H]
\figurenum{4}
\plotone{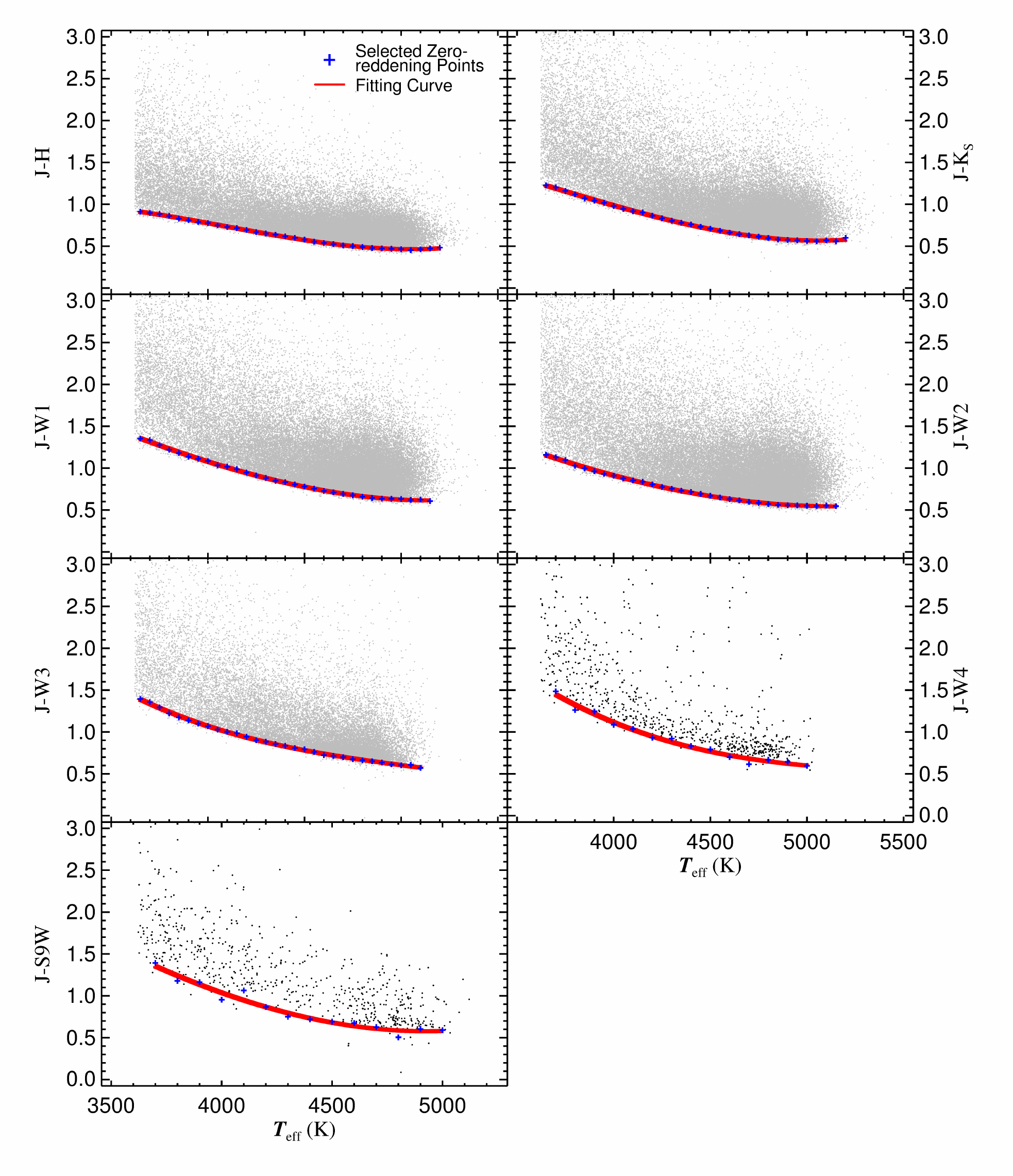}
\caption{The same as Fig.~\ref{fig:ic-d} but for giant stars from the APOGEE survey. }
\label{fig:ic-g}
\end{figure}

\begin{figure}[H]
\figurenum{5}
\plotone{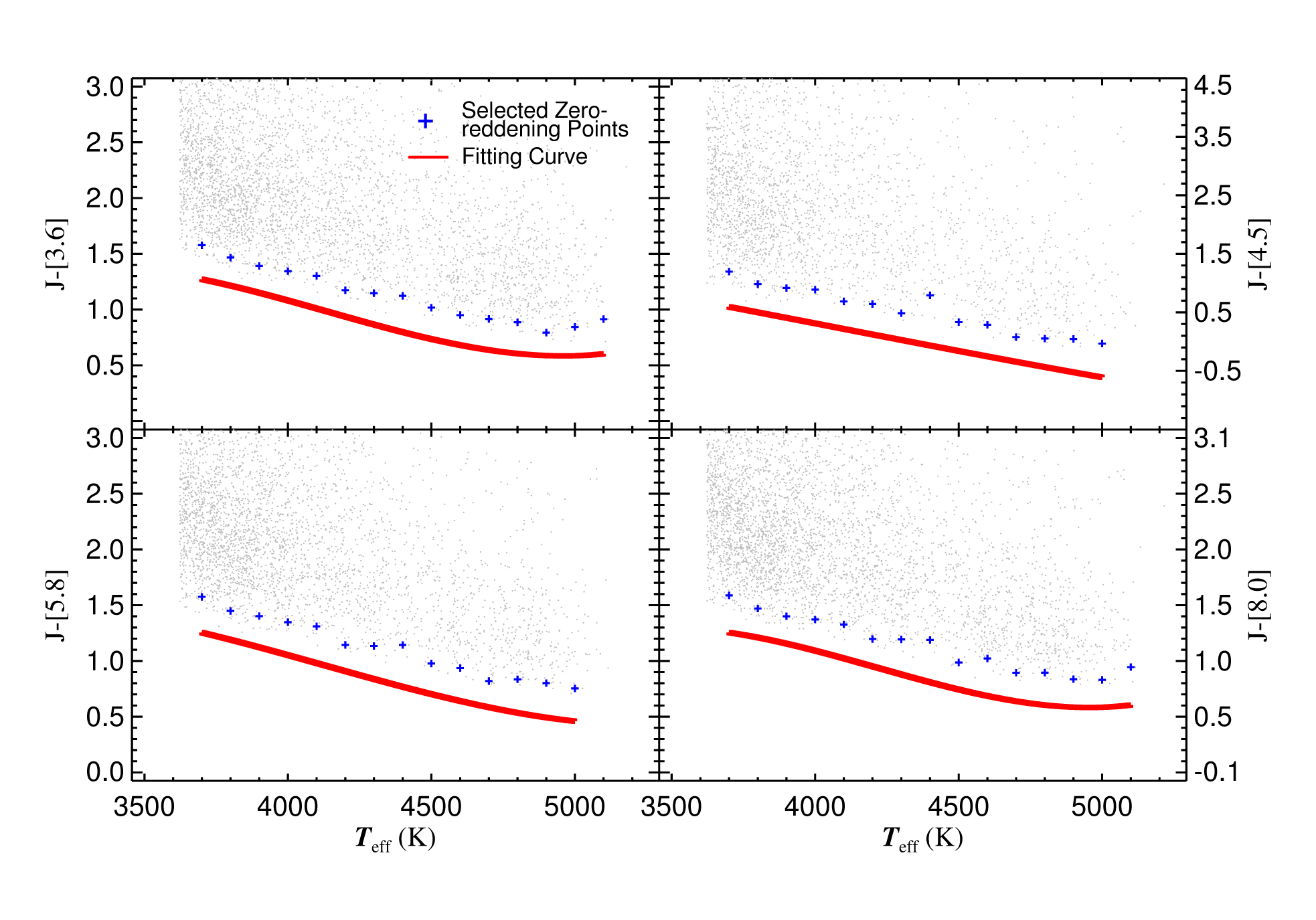}
\caption{The same as Fig.~\ref{fig:ic-g} but for \emph{Spitzer}/IRAC bands. Intrinsic color and function are shifted to compensate for the interstellar reddening, and the amount of shift can be found in Sec.~\ref{sec:method}.}
\label{fig:ic-gi}
\end{figure}

An example result of the selection can be found in Fig.~\ref{fig:ic-d}, where the blue crosses are selected to represent the intrinsic color indexes in a given bin of $\Teff$ according to the rules described above. With these discrete intrinsic color indexes determined, a third order polynomial function is fitted of the intrinsic color index $C_{\lambda1\lambda2}^{0}$ between bands $\lambda1$ and $\lambda2$  to the effective temperature $\Teff$:
\begin{equation}
C_{\lambda1\lambda2}^{0} = a_0 + a_1 \times \Teff + a_2 \times \Teff^2 + a_3 \times \Teff^3.
\label{eq:icr}
\end{equation}
This function defines the blue edge of the $J-\lambda$ vs. $\Teff$ diagram. Figs.~\ref{fig:ic-d}-\ref{fig:ic-gi} show the results of fitting in all the color indexes in study. Function form, either exponential or quadratic, is tested as well, and no apparent difference appears in the selected range of effective temperature. As this function is mathematical instead of physical, the form takes no effect in the result, while no extrapolation should be taken to lower or higher effective temperature.

The method works very well for the bands in all-sky surveys, i.e. 2MASS, \emph{WISE} and \emph{AKARI}. In the case of \emph{Spitzer}/IRAC bands, some modification must be taken, which is already pointed out by \citet{Xue2016}. The \emph{Spitzer}/GLIMPSE program targeted the Galactic plane within $|b| < 5 \degr$, where the interstellar extinction is almost unavoidable. This low-latitude area invalidates the condition of the method that the sample does contain zero-extinction star so that even the bluest star in the sample experiences some extinction and its color is not intrinsic. Based on the analysis and result of \citet[][Table 4]{Xue2016}, we calculated the difference $\Delta C_{\rm JK_{S}}^{0}$ in $C_{\rm JK_{S}}^{0}$  from all stars with that from the stars observed in the IRAC bands, which corresponds to the color excess $E_{\rm JK_{S}}$ of the bluest stars in the GLIMPSE survey. Then the color excess in band $\lambda$, $E_\mathrm{J\lambda}$  is calculated through $E_{\rm JK_{S}}$ and the ratio $E_{\rm K_{S}\lambda}/E_{\rm JK_{S}}$:
\begin{equation}
\label{eq:e}
E_\mathrm{J\lambda} =  E_\mathrm{JK_s}-E_\mathrm{K_s\lambda} = (1 + \frac{E_\mathrm{K_s\lambda}}{E_\mathrm{JK_s}})E_\mathrm{JK_s}
\end{equation}
It is found that $\Delta C_{\rm JK_{S}}^{0} =E_\mathrm{JK_s}$ is 0.0395 for dwarf stars and 0.216 for giant stars, so a shift of -0.050 (bluewards) for $J-[3.6]$ and a shift of -0.052 for $J-[4.5]$ are adopted for dwarf; a shift of -0.27 for $J-[3.6]$, -0.28 for $J-[4.5]$, -0.29 for $J-[5.8]$, -0.28 for $J-[8.0]$ is adopted for giants.

\begin{figure}[H]
\figurenum{6}
\plotone{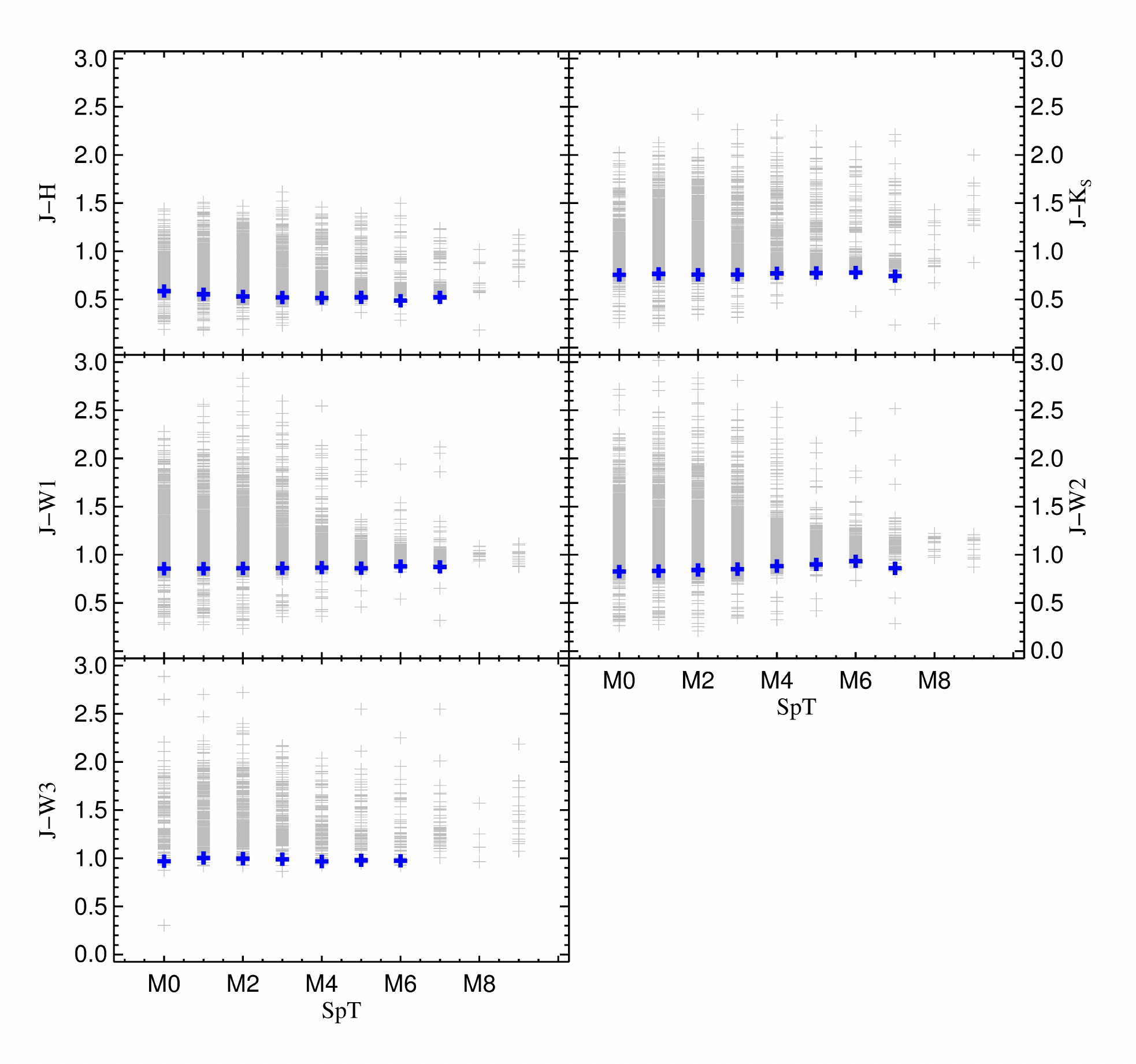}
\caption{The color -- subtype diagram of M-type dwarfs, grey crosses for photometrically selected stars, blue crosses for the median value of the bluest 5\% stars.}
\label{fig:ic-md}
\end{figure}

\begin{figure}[H]
\figurenum{7}
\plotone{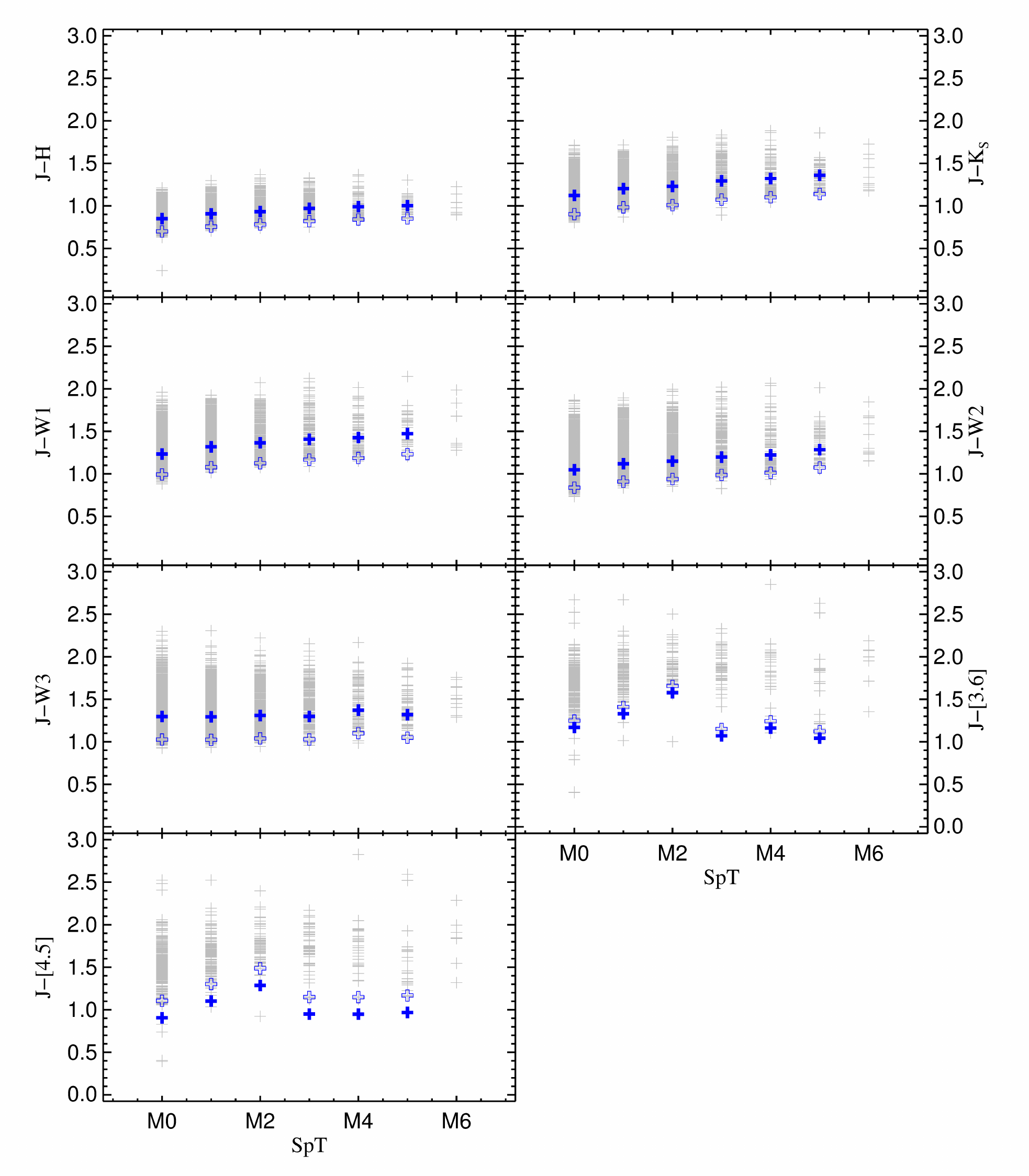}
\caption{The same as Fig.~\ref{fig:ic-md}, but for M-type giants; blue open cross for the median value of the bluest 5\% stars, and blue solid crosses for the intrinsic colors after calibrated by the APOGEE result, see Sec.~\textbf{\ref{sec:RG}}.}
\label{fig:ic-mg}
\end{figure}

As for M stars, the diagram of spectral type (SpT) vs. $J-\lambda$ replaces the $\Teff$ vs. $J-\lambda$ diagram because the adopted catalog lacks stellar parameters due to large uncertainty for such late-type stars. Stars are only classified into different spectral sub-types cursorily and the median value of 5\% bluest stars is set as preliminary intrinsic color. No fitting is applied and intrinsic color is presented at given spectral sub-types. Fig.~\ref{fig:ic-md} and~\ref{fig:ic-mg} show the results from this method for dwarfs and giants respectively.

\section{Result and Discussion}
\label{sec:result}

The results of three-order polynomial fitting to the intrinsic color indexes are shown in Table~\ref{tab:ic-dp} and \ref{tab:ic-gp} for dwarf and giant stars respectively. With these coefficients, the intrinsic color indexes can be calculated straightforward from $\Teff$ in the range specified in the last column of the tables. For convenience and comparison, the color indexes used for fitting at some typical $\Teff$ are listed in Table~\ref{tab:ic-dn} and \ref{tab:ic-gn}. The intrinsic color listed in \citet{Allen} provides a convenient way to compare. But the Allen's value is adopted from \citet{Bessell1988}, whose filters have distinct property with the 2MASS Survey. A transformation formula from \citet{Carpenter} is applied to the \citet{Bessell1988} value into the 2MASS magnitude system. The transformed result is marked as ``\citet{Bessell1988}'' on Fig.~\ref{fig:comp}, whose difference with no-transformation is around 0.0045 mag.

\subsection{The $\log{g}$ effect}

The effect of $\log{g}$ on giants' intrinsic colors is coupled with the effect of $\Teff$. Because the giant branch is nearly perpendicular to the $\log{g}$ lines in the H-R diagram, $\log{g}$ increases (from 1 to 3.0) with $\Teff$ monotonically, which means $\log{g}$ is not a fully independent parameter to affect the intrinsic color. To decompose the effect of $\log{g}$ from $\Teff$ is neither easy. The mean value of $\log{g}$ standard deviation on each $\Teff$ bin is only 0.22 that compares to  the error of $\log{g}$. For dwarfs, the effect of $\log{g}$ is not completely coupled with $\Teff$. However, the $\log{g}$ range of dwarf is narrower (from 3.7 to 5) than giant, while the error of $\log{g}$ from the LAMOST catalog is  around 0.5 (0.7 at most), which makes it difficult to further separate the intrinsic variation of $\log{g}$  from measurement uncertainty.  Therefore, our sample is roughly divided into two parts (giant and dwarf). Higher accuracy determination of  stellar parameters will make it possible to figure out the effect of $\log{g}$.

\subsection{Metallicity effect}

\begin{figure}[H]
\figurenum{8}
\plotone{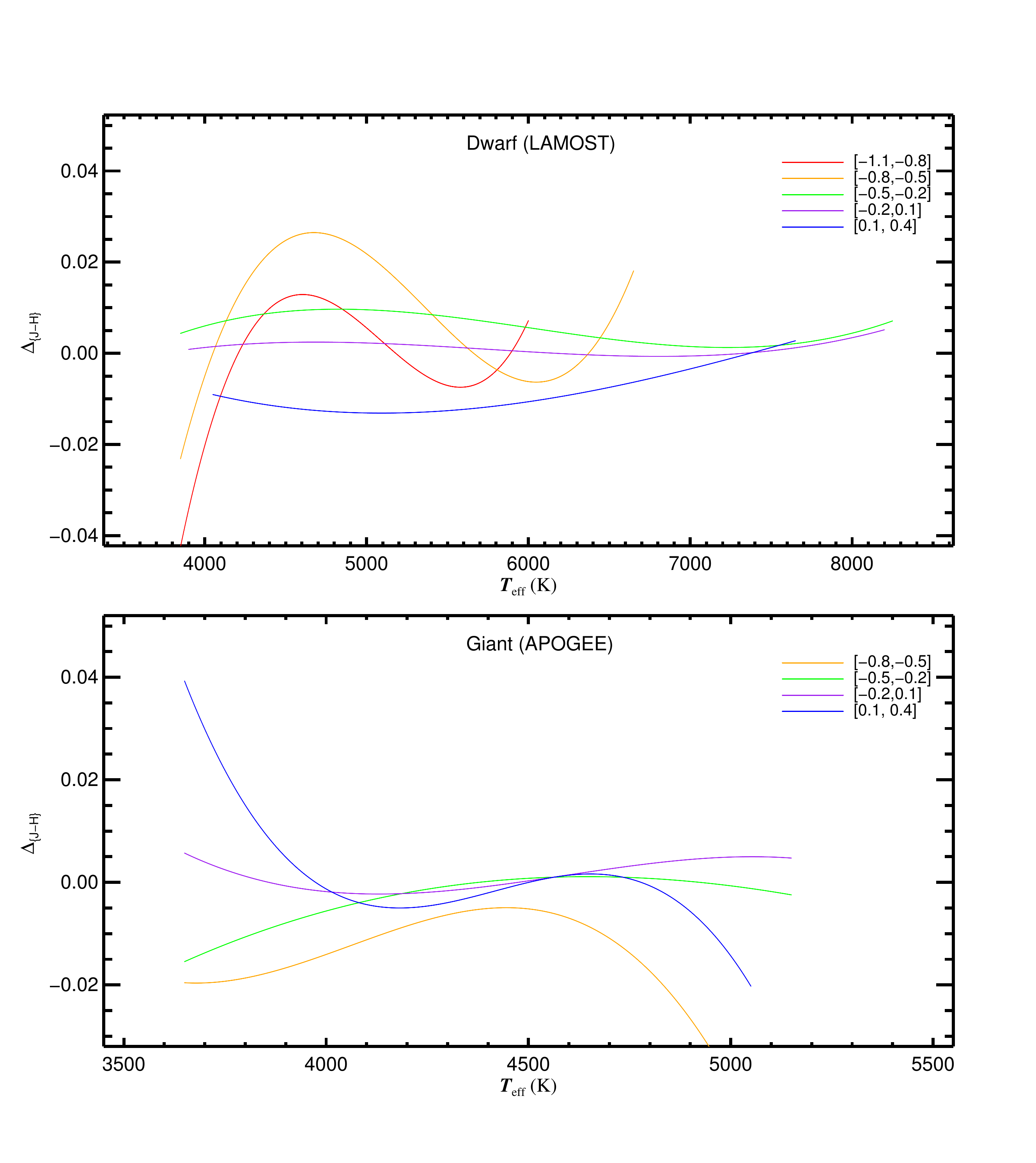}
\caption{The difference in $J-H$ of [Fe/H] bins with the metal-normal result.}
\label{fig:metalc}
\end{figure}

Fig.~\ref{fig:metalc} shows the difference in the intrinsic color $C_{\rm JH}^{0}$ for various metallicity ranges with  the metal-normal sample. It can be seen that metallicity has little effect on this color index. The difference is on the order of a couple of percentage magnitude, comparable to the photometric uncertainty. At the lower temperature end, the effect can be around 0.03 mag, otherwise less than 0.01 mag mostly. Considering that the stellar parameters at the lower temperature end suffers relatively large uncertainty, this phenomenon at low $\Teff$  should be treated cautiously, and may not be completely attributed to metallicity effect. In addition, [Fe/H] has smaller effect on dwarfs than giants. Except the metal-poor stars, all other stars have discrepancy smaller than 0.01 when compared with the whole metal-normal sample. As for giant, low [Fe/H] stars have the similar result with the whole metal-normal sample, and larger [Fe/H] value tends to increase intrinsic color at low $\Teff$ and decrease at high $\Teff$. However, the discrepancy is mostly within 0.02. Thus these discrepancy is considered as one of the uncertainty of intrinsic color.

The influence of metallicity on other colors is on the same order of magnitude. At low temperature, the metallicity effect can be as big as about 0.03 mag, which may not be fully attributable to metallicity as explained earlier. At $\Teff > 4000~\mathrm{K}$, the difference with the metal-poor sample is usually less than 0.02 mag.

\subsection{Dwarf}
\label{sec:RD-D}

The result from LAMOST is recommended for the dwarf stars, because APOGEE lacks a good sample of dwarfs and RAVE covers a smaller range of $\Teff$ than LAMOST. Though the edge is not very clear in the result of LAMOST (Fig.~\ref{fig:ic-d}), the large amount of stars ensures the intrinsic color still going along the edge. There is a bulge of stars at $\Teff$ from $\sim$5000\,K to $\sim$6500\,K below the intrinsic color line in all color indexes, which may come from the bias in template matching within this $\Teff$ range. The distribution of dwarfs in the $J-[3.6]/J-[4.5]$ vs $\Teff$ diagram is even more scattering but still exhibits a visible edge. Consequently, a rejection of stars beyond 3 sigma deviation from the mean value of a given $\Teff$ bin is not enough to exclude the outliers, so a 2-sigma rule of rejection replaces. The \emph{WISE}/$W4$ band has much fewer (806) sources which can be identified in the LAMOST survey, but there is a group of stars in the $J-W4$ diagram which clearly congregate near the blue envelope. A linear fitting of stars yields the relation: $C_{\rm JW4}^0 = -0.00023 \times \Teff + 1.89$.

The derived intrinsic color index $C_{\rm JH}^0$ agrees very well with classical values listed in \citet{Bessell1988}. Fig.~\ref{fig:comp} shows that the difference is mostly within 0.03 mag, and the average discrepancy is only 0.014. Even for the M-type  dwarfs, the tendency coincides. In the overlapping $\Teff$ range from about 4000\,K to 7000\,K, LAMOST shows no systematic difference with RAVE. The \citet{Ducati} result is discrepant from the other three ones, in particular at relatively lower $\Teff (< 5000$\,K).

\subsection{Giant}
\label{sec:RG}

The result of APOGEE is recommended for giant stars because of the large sample and clear edge in the $J-\lambda$ vs. $\Teff$ diagram that can be attributed to precise determination of stellar parameters.  Even the $J-W4$ vs. $\Teff$ diagram of metal-normal giants still has a clear blue edge although every bin of $\Teff$ has fewer than 100 stars. The results of four IRAC bands are also reliable with an unambiguous definition of the blue edge. In most bands, the color index $C_{\rm J\lambda}^{0}$ decreases with $\Teff$ as expected from a blackbody radiation that approximates the stellar radiation reasonably well in particular in the infrared bands. But at $\Teff = $5200\,K, a visible upwards tendency of $C_{\rm J\lambda}^{0}$ appears in a few bands, $C_{\rm JH}^{0}$, $C_{\rm JK_{S}}^{0}$, $C_{\rm J[3.6]}^{0}$, and $C_{\rm J[8.0]}^{0}$. For the IRAC [3.6] and [8.0] bands, there are not many sources in the bin of $\Teff$=5200\,K, such tendency may not be true.

\begin{figure}[H]
\figurenum{9}
\plotone{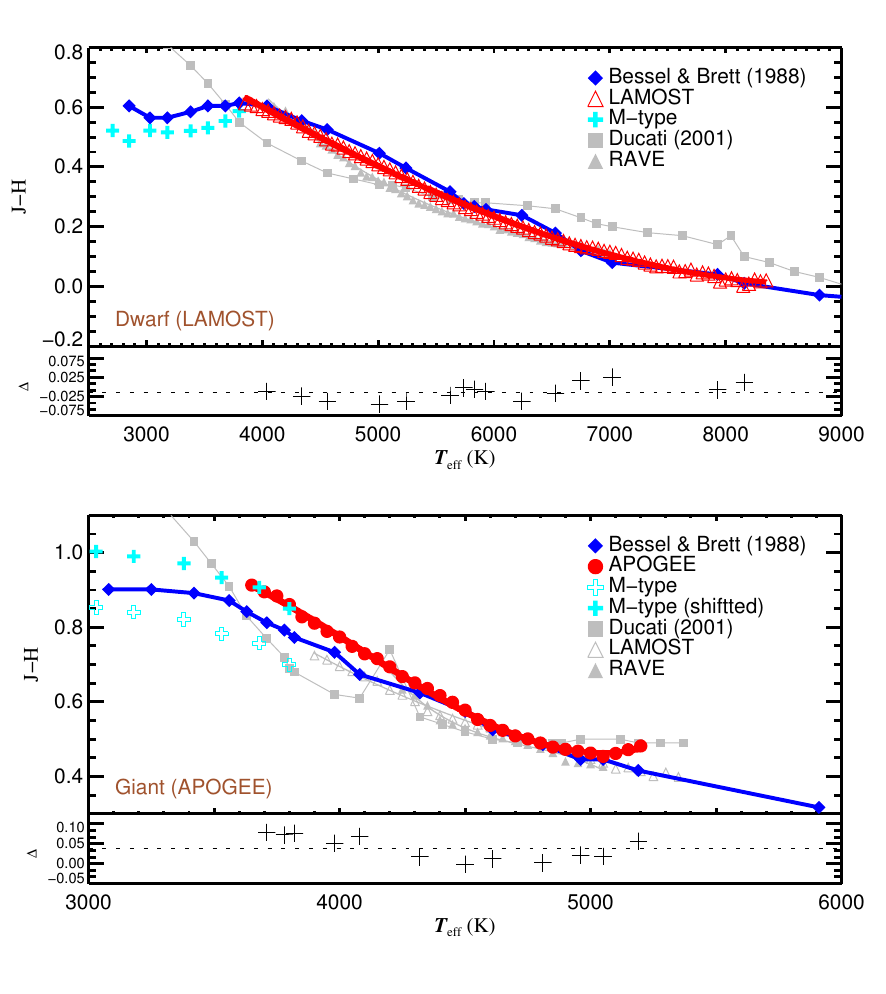}
\caption{Comparison of results from different spectroscopic surveys and previous results. $\Delta$ is the difference of this work and the result of \citet{Bessell1988}, and dashed line is the mean of $\Delta$. The cyan solid crosses are calibrated with the APOGEE result and shifted for original results about M-type stars, see Sec.~\ref{sec:method}.}
\label{fig:comp}
\end{figure}

The intrinsic color index $C_{\rm JH}^0$ derived from APOGEE matches that from \citet{Bessell1988} in the given range of $\Teff$ from 3650-5200\,K. The difference is mostly smaller than 0.05 mag with a mean of 0.039 (Fig.~\ref{fig:comp}). Such consistency is also found with the very recent determination of $C_{\rm JH}^0$ in a similar way by \citet{Xue2016}. The upwards turn at 5200\,K in the APOGEE result disappears in \citet{Bessell1988}, neither appears in the LAMOST result, while the RAVE sample does not extend to this high temperature. In fact, the LAMOST sample has a better defined blue edge at $\Teff >$ 4800\,K and should be a better indicator of $C_{\rm JH}^0$ there. So the upward tendency around 5200\,K in the APOGEE result is not reliable.

At the low $\Teff$ end, APOGEE cuts at about 3650\,K. For $\Teff <$ 3650\,K that is late-M-type giants, only the LAMOST results are available, which are much bluer than that from \citet{Bessell1988}, by an amount of about 0.2 mag as $\Teff$ from 3600\,K to about 3200\,K. The tendency of \citet{Ducati} result coincides with Bessell \& Brett's in this $\Teff$ range in that both becomes redder with decreasing $\Teff$. The discrepancy also exists for the M-type stars with the APOGEE result at the overlapping range of $\Teff$ from about 3600\,K to 3800\,K.  \citet{MTypeG} pointed out that there are about 4.7\% dwarfs in the sample of giant M-type stars. Although we tried to remove the dwarf contamination by applying the color criteria, the sample may not be pure of giant stars. It is apparent that our derived colors of M giants are between M dwarfs and giants. In order to keep internal consistency, the derived color indexes of M-type giants are shifted to match the analytical $J-\lambda$ vs. $\Teff$ relation derived from APOGEE at the overlapping range of $\Teff$, i.e. from about 3800\,K to 3600\,K and for subtypes M0 and M1. It turns out that the shifts are 0.15 for $J-H$, 0.22 for $J-\Ks$, 0.24 for $J-W1$, 0.21 for $J-W2$, 0.27 for $J-W3$, -0.08 for $J-[3.6]$, -0.2 for $J-[4.5]$ respectively. The consistency with APOGEE is reasonably good for all these color indexes except $J-[3.6]$, $J-[4.5]$ and $J-W3$ for which the results are not recommended. Table~\ref{tab:ic-mg} are the results after the shift. The results on M-type giants after calibrated by the APOGEE results agree very well with that from  \citet{Bessell1988}, which is shown in Fig.~\ref{fig:comp}. However, uncertainty of fitting result will affect the shift value, and the assumption that the M giant shift is the same for all temperatures may not be completely correct, so the colors at lower temperatures are more uncertain.

\subsection{Uncertainty}

The factors that contribute to the uncertainty of color index are: wrong match of star from photometric catalog to spectroscopic catalog, photometric errors in two bands of color index, error of  $\Teff$ and [Fe/H] caused by the flaw of pipeline, error of fitting the relation between color index and $\Teff$ points, the choice of bluest fraction, and the error of shift value for the IRAC bands. Because the accuracy of positional match is very high and the matched sample is large which forms the basis for our statistical method, the error from wrong match can be neglected. The mathematical fitting should neither affect the result much because the residual of fitting with a polynomial function is very small. The photometric and $\Teff$ value of the bluest fraction are re-calculated by 2000 times' Monte-Carlo method: each new value is the sum of catalog value and a Gaussian random number determined by its error; the result is shown in Table \ref{tab:mc}. The result is on the order of 0.01, some are less, and both smaller than the photometric quality control criteria. The effect of [Fe/H] and different bluest fraction will introduce an error of about 0.02. As for error of shift value, they can be derived through Equation~\ref{eq:e} if the shift is used to eliminate reddening. This kind of error is around 0.026, about half of IRAC bands quality control value. Shift error of M giant is difficult to estimate and have been discussed in Sec.~\ref{sec:RG}. The resultant uncertainty in observed color index equals to:
\begin{equation}
\sigma_\mathrm{C_{\rm J\lambda}} = \sqrt{\sigma_\mathrm{MC}^2+\sigma_\mathrm{[Fe/H]}^2+\sigma_\mathrm{ratio}^2+\sigma_\mathrm{shift}^2}
\end{equation}
where $\sigma_\mathrm{MC}$ refers to the result from  Monte-Carlo simulation, $\sigma_\mathrm{[Fe/H]}$ and $\sigma_\mathrm{ratio}$ refer to the error of various [Fe/H] and bluest fraction (they are both set as 0.02), and $\sigma_\mathrm{shift}$ refers to the error caused by shift (0.026 for de-reddening shift). The resultant uncertainty is presented in Table ~\ref{tab:uncer} and most uncertainties are around 0.03. However the absence of M giant shift error makes the error of M giant under-estimated.

\subsection{$H-W2$/$K-W1$}

The Rayleigh-Jeans approximation works better for stellar radiation at longer wavelength although specific wavelength depends on $\Teff$. \citet{Majewski2011} suggested that the intrinsic color involving near-infrared and mid-infrared bands may be constant. In particular, they recommended $C^0_\mathrm{H[4.5]}$ and $C^0_\mathrm{\Ks[3.6]}$ that vary in a very small range over a wide range of spectral type stars. This property makes it very convenient to estimate stellar color excess and interstellar extinction, which is then widely adopted. From the cross-identified data set, we check whether this color index changes in the range of $\Teff$ covered by the spectroscopic survey. Since the \emph{WISE}/W2 band  highly resembles the IRAC [4.5] band and \emph{WISE} is an all-sky survey, the W2 band is taken to replace the [4.5] band.  As can be seen from Fig.~\ref{fig:cons}, $C_{\rm HW2}^0$ remains almost constant at $\Teff >$ 4000\,K for giants and $\Teff >$ 5000\,K for dwarfs, and this constant value agrees very well with the recommended value of 0.08 by \citet{Majewski2011}. When $\Teff$ becomes smaller, $C_{\rm HW2}^0$ rises. The amount of increase is about 0.15 mag for dwarfs and 0.2 mag for giants. Such increase would bring the same amount deviation of color excess from assuming constant intrinsic colors. Fig.~\ref{fig:cons} tells that the intrinsic colors become redder at the lowest effective temperatures, and therefore using a constant value to determine reddening will over-estimate extinction for the coolest stars.

Another color index that spans a comparably narrow range as $C_{\rm HW2}^0$  is $C_{\rm \Ks W1}^0$, which can play an equivalent role in determining the interstellar extinction by the Rayleigh-Jeans Color Excess method \citep{Majewski2011}. Fig.~\ref{fig:cons} shows that this color index remains almost a constant of zero for dwarfs with 4000\,K $< \Teff <$ 8000\,K. For giants, there is a variation of about 0.05 mag around $C_\mathrm{\Ks W1}^0=$ 0.05 at $\Teff = $ 3600\,K. $C_\mathrm{\Ks W1}^0$ actually has less variation in comparison with $C_{\rm HW2}^0$, in particular at lower temperature for dwarf stars. Therefore, $C_{\rm \Ks W1}^0$ is a more accurate standard for calculating interstellar extinction with the RJCE method.

The range of $C_{\rm J\Ks}^0$ for both dwarf and giant stars was presented in Fig. 4 and 5 of \citet{Majewski2011}. In our study, the $\Teff$ range of dwarf (LAMOST) is 3850-8400 K, which corresponds to A4-M0 stars according to Table 7.6 of  \citet{Bessell1988}, and the corresponding $C_{\rm J\Ks}^0$ range is [0.048, 0.794], which is slightly wider than the range of \citet{Majewski2011} ([0.1, 0.7]). This small difference may be due to the absence of late type dwarf in Majewski's data set, but these two results coincide with each other. For the giant stars, this work yields a range of [0.578, 1.22] in $C_{\rm J\Ks}^0$ from APOGEE with $\Teff \in [3650, 5200]$\,K , while the range by Majewski is [0.85, 1.2]. Apparently, our smallest index is much bluer. However, if the red clump stars in the work of \citet{Majewski2011} is also counted as giant stars, then their range would extend to [0.55, 1.2] that is very consistent with ours. This comparison shows that there is a general consistency between our empirical method and stellar modelling method.

\section{Summary}
\label{sec:summary}

With the stellar parameters derived from large-scale spectroscopic surveys, the intrinsic color indexes are derived in the infrared bands of large-scale photometric surveys, which involve the 2MASS/$JH\Ks$, \textit{WISE}/W1-4, \textit{Spitzer}/IRAC1-4 and \textit{AKARI}/S9W bands. By fitting the relation of color index with effective temperature $\Teff$ of selected zero-reddening stars, an analytical relation is derived  between the intrinsic color index $C^0_\mathrm{J\lambda}$ and $\Teff$. This relation is convenient and accurate for calculating the intrinsic color index at a given $\Teff$. The color indexes used for the fitting of analytical relation are also presented (Table~\ref{tab:ic-dn}, \ref{tab:ic-gn}, \ref{tab:ic-md} and \ref{tab:ic-mg}). For the M-type stars, the intrinsic colors are derived at some sub-types instead of $\Teff$ due to shortage of accurate $\Teff$ for M-type stars. The uncertainty of each color indexes are derived. In addition, the intrinsic color indexes $C_{\rm HW2}^0$ and $C_{\rm KW1}^0$ are derived and their constancy is discussed.

In general, the tendency of our results agree with the classical results from  \citet{Bessell1988}, but with a systematic bluer color for dwarf stars.  Meanwhile, there is some discrepancy with the Ducati's result, in particular at relatively lower $\Teff$. Metallicity has little effect on these infrared colors.

{\bf{Acknowledgements}}

We thank Dr. Jian Gao, Shu Wang, and Mengyao Xue for very helpful discussion, and the anonymous referee for very constructive suggestions. This work is supported by China's NSFC projects 11533002, 11503002, 11373015, 973 Program 2014CB845702 and ``the Fundamental Research Funds for the Central Universities''. This work makes use of the data from the surveys by LAMOST, RAVE, SDSS/APOGEE, 2MASS, \emph{Spitzer}/GLIMPSE, \emph{WISE} and \emph{AKARI}.

%
%

\clearpage

\newpage

\newpage

\newpage

\newpage

\newpage

\newpage
	
\newpage

\newpage

\newpage

\begin{figure}[H]
\figurenum{10}
\plotone{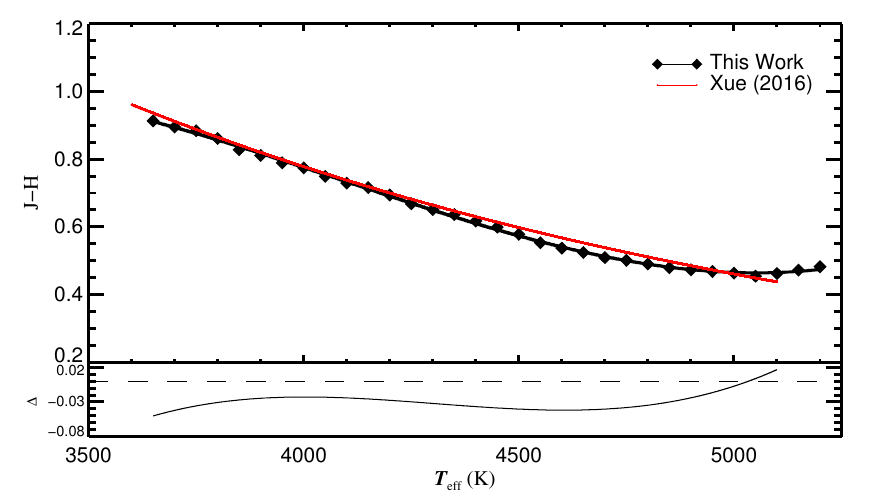}
\caption{Comparison of our result with that in \citet{Xue2016}. $\Delta$ is the subtraction of our result and \citet{Xue2016}, and dashed line indicates the zero value.}
\label{fig:xue}
\end{figure}
\newpage

\begin{figure}[H]
\figurenum{11}
\plotone{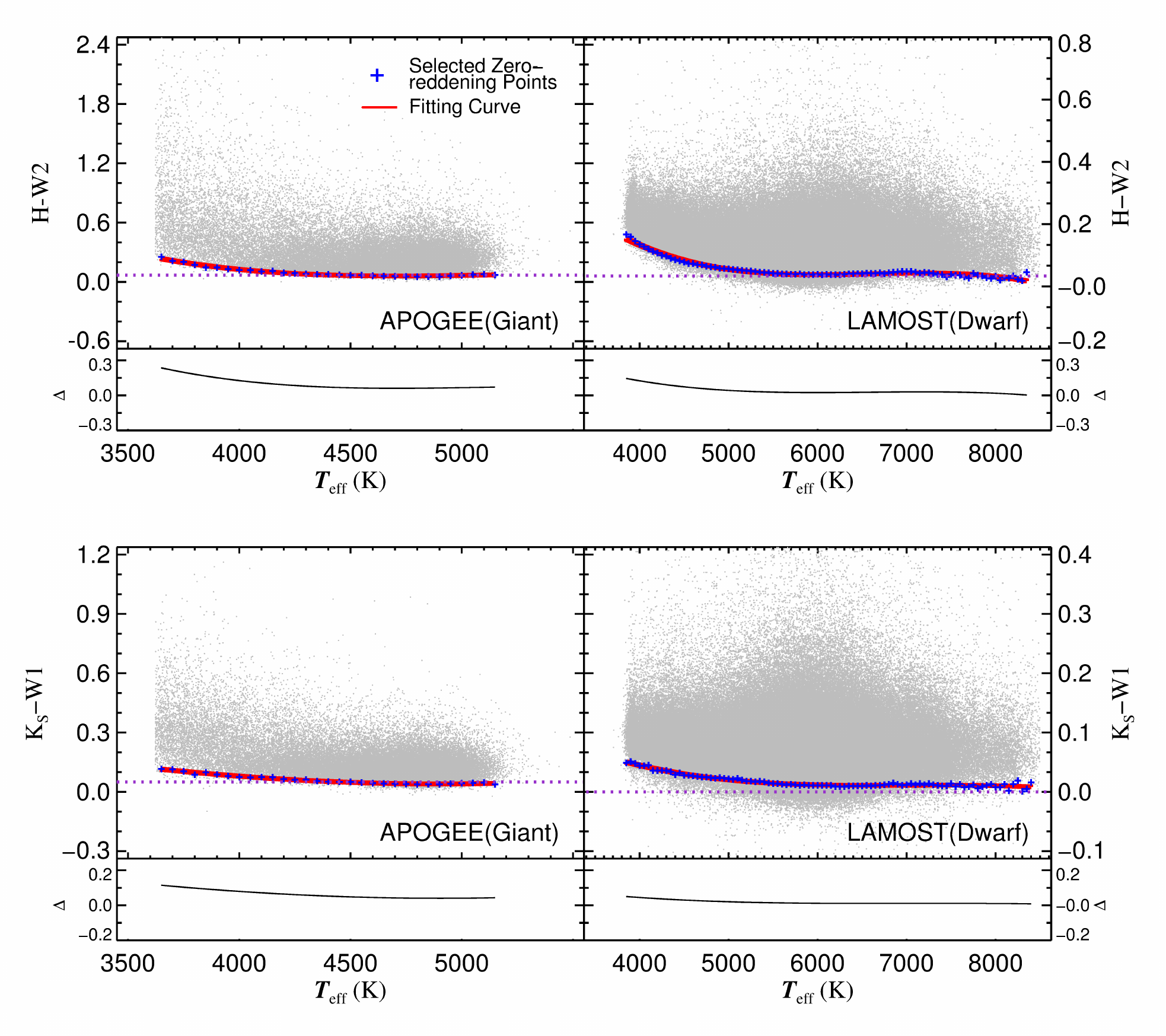}
\caption{Variation of $H-W2$ and $K-W1$ with $\Teff$. The brown dashed line denotes the constant suggested by \citet{Majewski2011}. $\Delta$ is the difference with the constants.}
\label{fig:cons}
\end{figure}
\newpage

\begin{table}[H]
\centering
\caption{Quality control for spectroscopic data.}
\begin{tabular}{c c c c c c}
\hline\hline
& $e_{\log{g}}$ & $e_\mathrm{[Fe/H]}$ & $e_{\Teff}$ & S/R & Other\\\hline
RAVE & $<$0.2 & $<$0.2 & $<$100\,K & $>$20 & Quality Flag$<$1\\
LAMOST & $<$0.7 & $<$0.3 & $<$300\,K & -- & Class: STAR\\
APOGEE & $<$0.22 & $<$0.1 & $<$200\,K & $>$100 & -- \\
\hline\hline
\end{tabular}
\label{tab:spec-restrict}
\end{table}

\begin{table}[H]
\centering
\caption{Quality control for photometric data.
}
\begin{tabular}{c c c c c c c}
\hline\hline
error & $e_\mathrm{J,H,K_s}$ & $e_\mathrm{W1,W2}$ & $e_{W3}$ & $e_{W4}$ & $e_{[3.6]-[8.0]}$ & $e_\mathrm{S9W}$\\\hline
value & $<0.03$ & $<0.03$ & $<0.05$ & $<0.1$ & $<0.05$ & $<0.1$\\\hline
\end{tabular}
\label{tab:phot-restrict}
\end{table}

\begin{table}[H]
\centering
\caption{Number of cross-identified stars that passed both spectroscopic and photometric data quality control.}
\begin{tabular}{c c c c c}
\hline\hline
 & 2MASS & \emph{WISE} & GLIMPSE & \emph{AKARI}\\
\hline
RAVE & 212,534 & 260,147 & 356 & 16,516\\
LAMOST & 1,130,775 & 1,130,775 & 23,808 & 844\\
APOGEE & 89,348 & 76,983 & 6,921 & 1,624\\
M-type Dwarf & 52,901 & 52,546 & 0 & 0\\
M-type Giant & 7,262 & 7,262 & 635 & 238\\
\hline\hline
\end{tabular}
\label{tab:number}
\end{table}

\begin{table}[H]
\centering
\caption{Number of four classes of stars.}
\begin{tabular}{c c c c c}
\hline\hline
& RAVE & LAMOST & APOGEE & M-Type\\\hline
metal-poor dwarf & 749 & 53,691 & 726 & \multirow{2}{*}{52,901}\\
metal-normal dwarf & 71,199 & 770,197 & 149 & \\
metal-poor giant & 7,785 & 29701 & 9,563 & \multirow{2}{*}{7,262}\\
metal-normal giant & 93,380 & 186,196 & 66,939 &
\\\hline\hline
\end{tabular}
\label{tab:class}
\end{table}

\begin{table}[H]
\centering
\caption{Uncertainty of intrinsic color from photometric error. }
\begin{tabular}{c c c c c c c}
\hline\hline
 & $\sigma_\mathrm{J-H,K_s,W1,W2}$ & $\sigma_\mathrm{J-W3}$ & $\sigma_\mathrm{J-W4}$ & $\sigma_\mathrm{J-[3.6],[4.5]}$ & $\sigma_\mathrm{J-[5.8],[8.0]}$ & $\sigma_\mathrm{J-S9W}$\\
\hline
Dwarf & 0.042 & 0.058 & 0.10 & 0.058 & -- & -- \\
Giant & 0.042 & 0.058 & 0.10 & 0.058 & 0.058 & 0.10\\
M Type Dwarf & 0.042 & 0.058 & -- & -- & -- & -- \\
M Type Giant & 0.042 & 0.058 & -- & 0.058 & -- & --\\
\hline\hline
\end{tabular}
\label{tab:err}
\end{table}

\begin{table}[H]
\centering
\caption{Coefficients in Equation~\ref{eq:icr} describing its relation to effective temperature $\Teff$ for dwarfs from LAMOST.}
\begin{tabular}{ c c c c c c }
\hline\hline
Intrinsic Color & $a_0$ & $a_1$ & $a_2$ & $a_3$ & $\Teff$ Range\\\hline
$C^0_\mathrm{JH}$ &  1.50e+00 &-2.13e-04 &-9.39e-09 & 1.63e-12 & 3850-8350\,K\\
$C^0_\mathrm{JK}$ &  2.35e+00 &-5.16e-04 & 2.88e-08 &-8.80e-15 & 3850-8400\,K\\
$C^0_\mathrm{JW1}$ &  2.73e+00 &-6.44e-04 & 4.64e-08 &-8.48e-13 & 3850-8400\,K\\
$C^0_\mathrm{JW2}$ &  3.41e+00 &-1.07e-03 & 1.22e-07 &-5.05e-12 & 3850-8400\,K\\
$C^0_\mathrm{JW3}$ &  3.98e+00 &-1.28e-03 & 1.47e-07 &-6.05e-12 & 3850-8150\,K\\
$C^0_\mathrm{JW4}*$ &  1.89e+00 &-2.30e-04 & -- & -- & 3950-7750\,K\\
$C^0_\mathrm{J[3.6]}$ &  3.55e+00 &-1.02e-03 & 1.01e-07 &-3.14e-12 & 3900-8250\,K\\
$C^0_\mathrm{J[4.5]}$ &  3.80e+00 &-1.18e-03 & 1.29e-07 &-4.70e-12 & 3900-8250\,K\\
\hline\hline
\end{tabular}
\label{tab:ic-dp}
\end{table}

\begin{table}[H]
\centering
\caption{Intrinsic color indexes of dwarfs adopted for fitting at typical temperatures.}
\begin{tabular}{ c c c c c c c c }
\hline\hline
$\Teff$\,(K) & $C^0_\mathrm{JH}$ & $C^0_\mathrm{JK}$ & $C^0_\mathrm{JW1}$ & $C^0_\mathrm{JW2}$ & $C^0_\mathrm{JW3}$ & $C^0_\mathrm{J[3.6]}$ & $C^0_\mathrm{J[4.5]}$\\\hline
3900 &  0.61 &  0.76 &  0.85 &  0.80 &  0.84 &  0.91 &  0.92\\
4000 &  0.59 &  0.74 &  0.83 &  0.77 &  0.83 &  0.83 &  0.86\\
4500 &  0.50 &  0.62 &  0.69 &  0.61 &  0.63 &  0.69 &  0.66\\
5000 &  0.41 &  0.50 &  0.57 &  0.50 &  0.52 &  0.56 &  0.55\\
5500 &  0.31 &  0.38 &  0.44 &  0.38 &  0.38 &  0.43 &  0.43\\
6000 &  0.23 &  0.29 &  0.34 &  0.30 &  0.27 &  0.34 &  0.35\\
6500 &  0.16 &  0.21 &  0.26 &  0.23 &  0.19 &  0.27 &  0.31\\
7000 &  0.12 &  0.16 &  0.21 &  0.19 &  0.16 &  0.26 &  0.28\\
7500 &  0.07 &  0.11 &  0.15 &  0.13 &  0.09 &  0.17 &  0.22\\
8000 &  0.02 &  0.06 &  0.10 &  0.07 &  --  &  0.21 &  0.25\\
8350 &  0.02 &  0.06 &  0.09 &  0.06 &  --  &  --  & --  \\
\hline\hline
\end{tabular}
\label{tab:ic-dn}
\end{table}

\begin{table}[H]
\centering
\caption{Coefficients in Equation~\ref{eq:icr} describing its relation to effective temperature $\Teff$ for giants from APOGEE.}
\begin{tabular}{ c c c c c c }
\hline\hline
Intrinsic Color & $a_0$ & $a_1$ & $a_2$ & $a_3$ & $\Teff$ Range\\\hline
$C^0_\mathrm{JH}$ & -8.13e+00 & 7.35e-03 &-1.90e-06 & 1.55e-10 & 3650-5200\,K\\
$C^0_\mathrm{JK}$ & -2.42e+00 & 4.32e-03 &-1.36e-06 & 1.23e-10 & 3650-5200\,K\\
$C^0_\mathrm{JW1}$ &  3.74e+00 & 3.06e-04 &-4.71e-07 & 5.71e-11 & 3650-5150\,K\\
$C^0_\mathrm{JW2}$ &  5.70e+00 &-1.44e-03 &-2.63e-08 & 2.15e-11 & 3650-5150\,K\\
$C^0_\mathrm{JW3}$ &  1.86e+01 &-1.00e-02 & 1.91e-06 &-1.25e-10 & 3650-5100\,K\\
$C^0_\mathrm{JW4}$ &  2.02e+01 &-1.03e-02 & 1.81e-06 &-1.07e-10 & 3700-5000\,K\\
$C^0_\mathrm{J[3.6]}$ & -2.13e+01 & 1.76e-02 &-4.43e-06 & 3.57e-10 & 3700-5100\,K\\
$C^0_\mathrm{J[4.5]}$ &  1.70e+00 & 3.62e-04 &-2.11e-07 & 1.72e-11 & 3700-5000\,K\\
$C^0_\mathrm{J[5.8]}$ & -9.49e+00 & 9.04e-03 &-2.37e-06 & 1.91e-10 & 3700-5000\,K\\
$C^0_\mathrm{J[8.0]}$ & -2.91e+01 & 2.28e-02 &-5.59e-06 & 4.42e-10 & 3700-5100\,K\\
$C^0_\mathrm{JS9W}$ &  7.31e+00 &-1.16e-03 &-3.60e-07 & 6.47e-11 & 3700-5000\,K\\
\hline\hline
\end{tabular}
\label{tab:ic-gp}
\end{table}

\begin{table}[H]
\centering
\caption{Intrinsic color indexes of giants adopted for fitting at typical temperatures.}
\begin{tabular}{ c c c c c c c }
\hline\hline
$\Teff$\,(K) & $C^0_\mathrm{JH}$ & $C^0_\mathrm{JK}$ & $C^0_\mathrm{JW1}$ & $C^0_\mathrm{JW2}$ & $C^0_\mathrm{JW3}$ & $C^0_\mathrm{JW4}$ \\\hline
3700 &  0.89 &  1.20 &  1.34 &  1.13 &  1.36 &  1.49\\
4000 &  0.77 &  0.98 &  1.09 &  0.91 &  1.07 &  1.09\\
4500 &  0.58 &  0.71 &  0.78 &  0.67 &  0.79 &  0.79\\
5000 &  0.46 &  0.56 &  0.63 &  0.55 &  0.60 &  0.60\\
5200 &  0.48 &  0.60 &  --  &  --  &  --  & --  \\
\hline
$\Teff$\,(K) & $C^0_\mathrm{J[3.6]}$ & $C^0_\mathrm{J[4.5]}$ & $C^0_\mathrm{J[5.8]}$ & $C^0_\mathrm{J[8.0]}$ & $C^0_\mathrm{JS9W}$ &\\\hline
3700 &  1.30 &  1.06 &  1.28 &  1.30 &  1.39 & \\
4000 &  1.07 &  0.90 &  1.06 &  1.08 &  0.95 & \\
4500 &  0.74 &  0.60 &  0.68 &  0.70 &  0.69 & \\
5000 &  0.57 &  0.41 &  0.46 &  0.54 &  0.59 & \\
5200 &  --  &  --  &  --  &  --  &  --  & \\
\hline\hline
\end{tabular}
\label{tab:ic-gn}
\end{table}

\begin{table}[H]
\centering
\caption{Intrinsic color indexes of M-type dwarf.}
\begin{tabular}{ c c c c c c }
\hline\hline
SpT & $C^0_\mathrm{JH}$ & $C^0_\mathrm{JK}$ & $C^0_\mathrm{JW1}$ & $C^0_\mathrm{JW2}$ & $C^0_\mathrm{JW3}$\\\hline
 M0 &  0.59 &  0.76 &  0.86 &  0.83 &  0.97\\
 M1 &  0.55 &  0.77 &  0.86 &  0.83 &  1.00\\
 M2 &  0.53 &  0.76 &  0.86 &  0.84 &  1.00\\
 M3 &  0.52 &  0.76 &  0.86 &  0.85 &  0.99\\
 M4 &  0.52 &  0.77 &  0.87 &  0.88 &  0.97\\
 M5 &  0.52 &  0.77 &  0.86 &  0.90 &  0.98\\
 M6 &  0.49 &  0.78 &  0.88 &  0.93 &  0.97\\
 M7 &  0.52 &  0.74 &  0.87 &  0.86 &  -- \\
\hline\hline
\end{tabular}
\label{tab:ic-md}
\end{table}

\begin{table}[H]
\centering
\caption{Intrinsic color indexes of M-type giant.}
\begin{tabular}{ c c c c c c c c}
\hline\hline
SpT & $C^0_\mathrm{JH}$ & $C^0_\mathrm{JK}$ & $C^0_\mathrm{JW1}$ & $C^0_\mathrm{JW2}$ & $C^0_\mathrm{JW3}$ & $C^0_\mathrm{J[3.6]}$ & $C^0_\mathrm{J[4.5]}$\\\hline
 M0 &  0.85 &  1.12 &  1.23 &  1.05 &  1.30 &  1.17 &  0.91\\
 M1 &  0.91 &  1.20 &  1.32 &  1.12 &  1.29 &  1.33 &  1.10\\
 M2 &  0.93 &  1.23 &  1.37 &  1.15 &  1.31 &  1.58 &  1.29\\
 M3 &  0.97 &  1.29 &  1.41 &  1.20 &  1.30 &  1.07 &  0.95\\
 M4 &  0.99 &  1.32 &  1.43 &  1.22 &  1.37 &  1.16 &  0.95\\
 M5 &  1.00 &  1.36 &  1.47 &  1.28 &  1.32 &  1.04 &  0.97\\
\hline\hline
\end{tabular}
\label{tab:ic-mg}
\end{table}

\begin{table}[H]
\centering
\caption{Result of Monte-Carlo simulation}
\begin{tabular}{ c c c c c c c}
\hline\hline
 & $J-H$ & $J-K_S$ & $J-W1$ & $J-W2$ & $J-W3$ & $J-W4$\\
\hline
LAMOST & 0.0011 & 0.0010 & 0.0011 & 0.0012 & 0.010 & -\\
APPGEE & 0.0035 & 0.0044 & 0.0054 & 0.0048 & 0.0078 & 0.043\\
M Type Dwarf & 0.0078 & 0.0067 & 0.0072 & 0.0078 & 0.0094 & -\\
M Type Giant & 0.012 & 0.0092 & 0.012 & 0.010 & 0.013 & -\\
\hline
 & $J-S9W$ & $J-[3.6]$ & $J-[4.5]$ & $J-[5.8]$ & $J-[8.0]$ & -\\
\hline
LAMOST & - & 0.0084 & 0.0076 & - & - & -\\
APPGEE & 0.051 & 0.014 & 0.017 & 0.015 & 0.013 & -\\
M Type Dwarf & - & - & - & - & - & -\\
M Type Giant & - & 0.023 & 0.020 & - & - & -\\
\hline\hline
\end{tabular}
\label{tab:mc}
\end{table}

\begin{table}[H]
\centering
\caption{\textbf{Uncertainty of color index}}
\begin{tabular}{ c c c c c c c}
\hline\hline
 & $J-H$ & $J-K_S$ & $J-W1$ & $J-W2$ & $J-W3$ & $J-W4$\\
\hline
LAMOST & 0.028 & 0.028 & 0.028 & 0.028 & 0.030 & -\\
APPGEE & 0.028 & 0.029 & 0.029 & 0.029 & 0.029 & 0.051\\
M Type Dwarf & 0.029 & 0.029 & 0.029 & 0.029 & 0.030 & -\\
M Type Giant & 0.031 & 0.030 & 0.031 & 0.030 & 0.031 & -\\
\hline
 & $J-S9W$ & $J-[3.6]$ & $J-[4.5]$ & $J-[5.8]$ & $J-[8.0]$ & -\\
\hline
LAMOST & - & 0.040 & 0.039 & - & - & -\\
APPGEE & 0.058 & 0.041 & 0.042 & 0.041 & 0.040 & -\\
M Type Dwarf & - & - & - & - & - & -\\
M Type Giant & - & 0.036 & 0.035 & - & - & -\\
\hline\hline
\end{tabular}
\label{tab:uncer}
\end{table}

\end{document}